\documentclass[sn-mathphys-num]{sn-jnl}

\usepackage{graphicx}%
\usepackage{multirow}%
\usepackage{amsmath,amssymb,amsfonts}%
\usepackage{amsthm}%
\usepackage{mathrsfs}%
\usepackage[title]{appendix}%
\usepackage{xcolor}%
\usepackage{textcomp}%
\usepackage{manyfoot}%
\usepackage{booktabs}%
\usepackage{algorithm}%
\usepackage{algorithmicx}%
\usepackage{algpseudocode}%
\usepackage{listings}%
\usepackage{natbib}%
\usepackage{lineno}%
\usepackage{comment}%

\newcommand*\patchAmsMathEnvironmentForLineno[1]{%
  \expandafter\let\csname old#1\expandafter\endcsname\csname #1\endcsname
  \expandafter\let\csname oldend#1\expandafter\endcsname\csname end#1\endcsname
  \renewenvironment{#1}%
     {\linenomath\csname old#1\endcsname}%
     {\csname oldend#1\endcsname\endlinenomath}}%
\newcommand*\patchBothAmsMathEnvironmentsForLineno[1]{%
  \patchAmsMathEnvironmentForLineno{#1}%
  \patchAmsMathEnvironmentForLineno{#1*}}%
\AtBeginDocument{%
\patchBothAmsMathEnvironmentsForLineno{equation}%
\patchBothAmsMathEnvironmentsForLineno{align}%
\patchBothAmsMathEnvironmentsForLineno{flalign}%
\patchBothAmsMathEnvironmentsForLineno{alignat}%
\patchBothAmsMathEnvironmentsForLineno{gather}%
\patchBothAmsMathEnvironmentsForLineno{multline}%
}

\theoremstyle{thmstyleone}%
\theoremstyle{thmstyletwo}%

\theoremstyle{thmstylethree}%

\begin{document}

\title[Article Title]{A general theory of quantum measurements}

\author{\fnm{Shizhong} \sur{Mei}} \email{mei.shizhong@gmail.com}

\affil{\orgdiv{} \orgname{} \orgaddress{\street{4941 E Sagewood Dr}, \city{Boise}, \postcode{83716}, \state{Idaho}, \country{U.S.A}}}

\abstract{A universal energy eigenvalue equation is proposed in this paper. It is proven that the unique set of eigenfunctions or preferred basis exists for any non-isolated sub-system. Applying the new eigenvalue equation to the relative motion of a hydrogen atom together with the derived relativistic Hamiltonian to quantify the impact of finite proton mass to the fine structure, correction to the fine structure is obtained as a result of the entanglement of the relative motion and the center-of-mass motion, which can be used to verify the correctness of the proposed eigenvalue equation. Applying the equation to the measurement of electron double-slit interference, it is analyzed that the photon packets with Lorentzian spectral lineshape, the  domain and state density of the sub-system, and the energy of the incident electron together determines the spontaneous emission rate of the incident electron. Photons generated in this process excite electrons from the valence band to the conduction band of the detector. Corresponding to any emitted or absorbed photon, the sub-system is found to be uniquely determined by maximizing the transition rate. This new principle is valid for atoms too. Closed-form expressions are obtained for the transition rates and example numerical results show good correlation between calculation and the position measurement experiment. The discovered common mechanisms that determine the sub-systems, the preferred bases, and transition rates form the foundation of a new, general, and consistent theory of quantum measurement.    }

\keywords{quantum measurements, preferred basis, fine structure, photon size, transition rate.}

\maketitle

\section{Introduction}\label{sec:1}

One hundred years have passed since since the initial development of quantum mechanics. This mechanics has withstood rigorous tests and is considered one of the most successful theory \cite{1}. The physical descriptions of quantum computation, quantum information \cite{2}, and quantum materials \cite{3} are all founded on quantum mechanics. Yet to study the foundation of quantum mechanics is necessary \cite{4}, for better understanding the strength and weakness and exploring solutions to address the weakness. There are critical cases that have demonstrated prediction-to-measurement correlation with striking accuracy, but the theory ignores the impact of other components that exist in a real system and is too specific to incorporate those components.

The calculation of Hydrogen atom spectrum is such an example. The non-relativistic Hamiltonian defines an energy eigenvalue equation. The closed-form solutions can be obtained by separating the center-of-mass motion and the relative motion influenced by the Coulomb potential. The center-of-mass moves freely in space so has continuous eigenvalues. The relative motion is equivalent to that of an electron with reduced mass constrained by the Coulomb potential of the nucleus fixed at one point. This motion has discrete eigenvalues for bound eigenfunctions. In the conventional method, the difference of the discrete eigenvalues has been the prediction for the energy of the spectral lines in quantum mechanics. It can be verified that the prediction matches the measurements \cite{5} very well.  

That prediction however suffers from a major weaknesses, i.e., the impact of other objects, albeit very weak, can not be incorporated into the calculation. Similar issue exists in the relativistic version \cite{6} that was obtained from Dirac equation \cite{7}. In that version, the mass of the proton was assumed infinite. The impact of a finite mass proton and other components in the surroundings are all ignored and can not be handled by the method. 

After careful analysis, the problem is found to be related to the form of energy eigenvalue equations. For an isolated object, the energy eigenvalue equation uses the Hamiltonian of the object as an operator to determine a set of eigenvalues and eigenfunctions for the object. However, there are no isolated objects in reality. The interaction between an object and its surroundings, no matter how weak it is, exists. The ideal case can be viewed as a model of the real system where the interaction terms approach zero. Due to continuity, the system Hamiltonian should be used for the ideal case. Consequently the system wave function should also be used. With these modifications, the old form of the energy eigenvalue equation of the object does not hold anymore. To maintain the eigenvalues and eigenfunctions, the form of equation needs to be changed. Applying the new equation to characterize the relative motion of atoms resolves the problem with the old methods.     

A new relativistic Hamiltonian of a hydrogen atom is derived in this paper. The assumed infinite mass of proton in the old method was for the purpose to obtain an equation for the relative motion of the electron. To avoid that weakness, the new derivation directly applies to the relative motion of the atom with relativistic form of kinetic energy. The new Hamiltonian contains the impact of finite mass proton and center-of-mass motion to the fine structure. The impact is quantified by applying the proposed energy eigenvalue equation. Although the deviation from the old results \cite{6} is small, e.g., $1.8\times10^{-6}eV$ total adjustment to $2P_{1/2}-1S_{1/2}$ energy gap for a hydrogen atom with $100eV$ kinetic energy of the center-of-mass motion, the new method is more comprehensive than the old one.  

To showcase the relevance of the proposed eigenvalue equation to the quantum measurement problem, the measurement of electron double-slit interference \cite{8} is reviewed. Applying the equation to a sub-system that describes the relative motion of the incident electron and part of the detector, components of bound states start to quickly increase when the wavefront of the incident electron approaches the maximum penetration depth. The interaction between charges and photons cause spontaneous emission. The interaction, the size of the photons, the total energy of the incident electron, and the sub-system within the domain of the photon together determine the transition rate. Closed form solutions are obtained and example numerical results show $2.8$ femto-seconds lifetime due to spontaneous emission and $1.1$ femto-seconds lifetime for all the emitted photons. It is analyzed that for any realized photon emission or absorption, the sub-system can be uniquely determined by maximizing the transition rate. This rule holds true for atoms too. The elevated conduction band electrons as the result of photon absorption will eventually fall back to the valence band and generate photons to indicate successful position measurement. Compared to the $1.2mm$ interference cycle of the interference pattern, those photons generated from the recombination of electrons and holes that are distributed in a sphere with $1.7um$ radius yield high resolution of position measurement.        

This explanation reveals the common new principles in different phenomena of atom spectrum generation and interference pattern detection. The revelation is significant, suggesting that all quantum measurements involving charges and photons can be described by the same general measurement theory.

The rest of the paper is organized as follows. Detailed explanation of the universal energy eigenvalue equation and basic analysis of the solution are provided in Section \ref{sec:2}. A new relativistic Hydrogen Hamiltonian is derived in section \ref{sec:3} together with the closed-form calculation of the impact of finite mass proton and the kinetic energy of the center-of-mass motion. Section \ref{sec:4} presents the derivation and numerical results to explain the measurement results of electron double-slit interference. The new principle to uniquely determine a sub-system is also analyzed in this section. Discussion and conclusion are provided in Section \ref{sec:5} and Section \ref{sec:6}, respectively.

\section{The universal energy eigenvalue equation}\label{sec:2}

Consider a system described by Hamiltonian $H$ and wave function $\mid\psi>$. The system can be viewed as the combination of a sub-system $s$ and its surroundings $s\_$. The system Hamiltonian $H$ consists of the Hamiltonian $H_s$ of the sub-system in isolation, the interaction term $H_{int}$ between the sub-system and its surroundings, and the Hamiltonian $H_{s\_}$ of the surroundings in isolation. According to quantum mechanics, a measurement of some observable $O_s$ of the sub-system will cause the system wave function to collapse. Right after the measurement, if the state of $s$ is an eigenfunctions $\mid\psi_{sj}>$ of the observable $O_s$, then the wave function $\mid\psi_{j}>$ of the system will be the normalized projection $\mid\psi_{sj}><\psi_{sj}\mid\psi>$.

When the interaction between $s$ and $s\_$ is weak enough to be ignored, the observable $O_s$ becomes $H_s$. In this situation, $s$ and $s\_$ have their respective wave functions $\mid\phi_s>$ and $\mid\phi_{s\_}>$. Atoms in the non-relativistic domain are examples of this situation where $s$ can be defined as a group of particles with relative motion while $s\_$ refers to the single particle with the center-of-mass motion. The Hamiltonian $H_s$ defines the energy eigenstates $\mid\phi_{sj}>$'s of the sub-system through energy eigenvalue equation 
\begin{linenomath*}
\begin{equation}
H_{s}\mid\phi_{sj}>=E_{sj}\mid\phi_{sj}>,\; j\in \mathbb{N} \label{eq:1}
\end{equation}
\end{linenomath*}
where $\mathbb{N}$ is a set of integers for $j$ such that $\mid\phi_{sj}>$'s represent a complete set of eigenfunctions. For convenience without loss of generality, all the eigenstates $\mid\phi_{sj}>$'s are assumed to be normalized and orthogonal to each other. Right after a measurement, if the sub-system wave function is $\mid\phi_{sj}>$, then the system wave function will be $\mid\phi_{sj}>\mid\phi_{s\_}>$.  

As pointed out earlier, the form of Eq.~\eqref{eq:1} does not meet the continuity requirement. The correct form should use $H$ instead of $H_s$ and the possible states immediately after a measurement are $\mid\phi_{sj}>\mid\phi_{s\_}>$'s instead of $\mid\phi_{sj}>$'s. The new equation should maintain the solution of Eq.~\eqref{eq:1}. This turns out to be easily satisfied if the new equation employs inner product. Due to the continuity requirement for the wave function, the more general wave function $\mid\psi_j>$ should be used instead of $\mid\phi_{sj}>\mid\phi_{s\_}>$ in the new equation. Now the same number of pairwise orthogonal and normalized functions $\mid\psi_j>$ as the number of eigenfunctions $\mid\phi_{sj}>$'s can be formed such that each $\mid\psi_j>$ is a linear combination of the set of $\mid\phi_{sl}>\mid\phi_{s\_}>$'s, i.e., $\sum\limits_{l}c_{jl}\mid\phi_{sl}>\mid\phi_{s\_}>$. By requiring $<\psi_j\mid H\mid\psi_k>$ to be nonzero only for $j=k$ in the new equation, the exact solution of equation \eqref{eq:1} can be obtained. With these considerations, the new equation can be written as
\begin{linenomath*}
\begin{equation}
\begin{aligned}
&<\psi_{sj}\mid\psi_{sk}>=\delta _{jk} , \\
&<\psi_j\mid H\mid\psi_k>=E_{j}\delta_ {jk},\;j,k\in \mathbb{N} \label{eq:2}
\end{aligned}
\end{equation}
\end{linenomath*}
where $\delta_{jk}$ is the Kronecker delta function and $E_{j}$ is the system energy corresponding to the wave function $\mid\psi_j>$. 

No new assumption has been made to obtain Eq.~\eqref{eq:2}. Since the Hamiltonian, wave function, and sub-system relevant to Eq.~\eqref{eq:2} are all generic, Eq.~\eqref{eq:2} is a universal eigenequation of observable and preferred basis. This equation defines the observable $O_s$ with eigenvalues $O_{sj}$'s equal to $E_{j}-<\psi_{j}\mid H_{s\_}\mid\psi_{j}>$'s and eigenfunctions $\mid\psi_{sj}>$'s, i.e.,
\begin{linenomath*}
\begin{equation}
O_s\mid\psi_{sj}>=O_{sj}\mid\psi_{sj}>, \; j\in \mathbb{N}. \label{eq:3}
\end{equation}
\end{linenomath*}
The observable $O_s$ consists of $H_s$ and an interaction term $\bar{H}_{int,s}$ that is a function of the sub-system and represents the effective interaction between the sub-system and its surroundings. The interaction term $\bar{H}_{int,s}$ satisfies 
\begin{linenomath*}
\begin{equation}
<\psi_{sj}\mid \bar{H}_{int,s}\mid\psi_{sk}> = <\psi_j\mid H_{int}\mid\psi_k>, \; j,k\in \mathbb{N}.
 \label{eq:4}
\end{equation}
\end{linenomath*}
Since $H_{int}$ is self-adjoint, it can be inferred from Eq.~\eqref{eq:4} that $H_{int,s}$ is also self-joint. This property plus $H_s$ being self-joint makes $O_s$ a self-adjoint operator. 

Equation ~\eqref{eq:2} indicates that the observable and preferred basis of a sub-system depend on the system Hamiltonian $H$ and wave function $\mid\psi>$. Given $H$ and $\mid\psi>$, the solution of Eq.~\eqref{eq:2} has to exist and be unique. Uniqueness is a strong requirement. It has not been met before generally.  
    
\subsection{Existence of general solution}\label{subsec:2.1}

When the interaction between $s$ and $s\_$  can not be ignored, the functions $\mid\psi_{sj}>$'s are not $\mid\phi_{sj}>$'s anymore but linear combinations of the latter, i.e.,
\begin{linenomath*} 
\begin{equation}
\mid\psi_{sj}>=\sum\limits_{l=1}^{\infty} a _{jl}\mid\phi_{sl}>, \; j\in \mathbb{N}. \label{eq:5}
\end{equation}
\end{linenomath*}
The corresponding inner product $<\psi_{sj}\mid\psi>$ equals $\sum\limits_{l=1}^{\infty}a_{jl}^*<\phi_{sl}\mid\psi>$ where the superscript * represents the complex conjugate. The first orthonormalization constraint in Eq.~\eqref{eq:2} requires
\begin{linenomath*}
\begin{equation}
\sum\limits_{l=1}^{\infty}a^* _{jl}\cdot a _{kl} = \delta_{jk}, \; j\leq k, \;k\in \mathbb{N}. \label{eq:6}
\end{equation}
\end{linenomath*}
Denote $h_{lmpq}$ the inner product $<\psi\mid\phi_{sl}><\phi_{sm}\mid H \mid\phi_{sq}><\phi_{sp}\mid\psi>$ and $C_{s\_l}$ the norm of $<\phi_{sl}\mid\psi>$. The second constraint in Eq.~\eqref{eq:2} can be written as
\begin{linenomath*}
\begin{equation}
\sum\limits_{l,m,p,q=1}^{\infty} a _{jl}a^* _{jm}h _{lmpq}a^* _{kp}a _{kq} = C_jE_j\delta_{jk} ,\; j\leq k, \;k\in \mathbb{N} \label{eq:7}
\end{equation}
\end{linenomath*}
where $C_j = \sum\limits_{l=1}^{\infty}a_{jl}^*a_{jl}C_{s\_l}$. Although the total number of unknowns in Eq.~\eqref{eq:6} and Eq.~\eqref{eq:7} equals the total number of constraints, the nonlinear terms in the equations make it difficult to solve them directly. An alternative is to break the calculation into multiple steps and apply linearization to obtain linear equations for the increments of the unknowns. In the limit that the number of steps approaches infinity, the error of linearization approaches zero so the exact solution can be obtained. 

More specifically, in each step, a small scale of $H_{int}$, i.e., $H_{int}/N$ with $N$ being a positive integer, is added to $H_s+H_{s\_}$. That means the combined Hamiltonian $H^{(n)}$ for the $n^{th}$ step equals $H_s+nH_{int}/N+H_{s\_}$. Replacing $H$, $\mid\psi_{sj}>$, $\mid\psi_{j}>$, and $E_j$ with $H^{(n)}$, $\mid\psi_{sj}^{(n)}>$, $\mid\psi_{j}^{(n)}>$, and $E_j^{(n)}$ respectively, Eq.~\eqref{eq:2} becomes 
\begin{linenomath*}
\begin{equation}
\begin{aligned}
&<\psi^{(n)}_{sj}\mid\psi^{(n)}_{sk}>=\delta _{jk} , \\
&<\psi^{(n)}_j\mid H^{(n)}\mid\psi^{(n)}_k>=E^{(n)}_{j}\delta_ {jk}, \; j,k\in \mathbb{N} \label{eq:8}
\end{aligned}
\end{equation}
\end{linenomath*}
where $\mid\psi_{j}^{(n)}>$ equals the normalized $\mid\psi_{sj}^{(n)}><\psi_{sj}^{(n)}\mid\psi>$ and $\mid\psi_{sj}^{(n)}>$ has similar expansion as Eq.~\eqref{eq:5} with coefficients $a_{jl}^{(n)}$'s. The constraints in Eq.~\eqref{eq:8} require 
\begin{linenomath*}
\begin{equation}
\sum\limits_{l=1}^{\infty}a^{(n)*} _{jl}\cdot a^{(n)} _{kl} = \delta_{jk}, \; j\leq k, \;k\in \mathbb{N} \label{eq:9}
\end{equation}
\end{linenomath*}
and
\begin{linenomath*}
\begin{equation}
\begin{aligned}
\sum\limits_{l,m,p,q=1}^{\infty} a^{(n)} _{jl}a^{(n)*} _{jm}h^{(n)} _{lmpq}a^{(n)*} _{kp}a^{(n)} _{kq} = C^{(n)}_jE^{(n)}_j\delta_{jk},\; \\
j\leq k, \;k\in \mathbb{N} \label{eq:10}
\end{aligned}
\end{equation}
\end{linenomath*}
where $h_{lmpq}^{(n)}$ represents the inner product $<\psi\mid\phi_{sl}><\phi_{sm}\mid H^{(n)} \mid\phi_{sq}><\phi_{sp}\mid\psi>$ and $C^{(n)}_j = \sum\limits_{l=1}^{\infty}a^{(n)}_{jl}*a^{(n)}_{jl}C_{s\_l}$.    

Initially $\psi_{sj}^{(0)}$ is the eigenfunction of $H_s$, $a_{jl}^{(0)}$ equals $\delta_{jl}$, and $E_{j}^{(0)}$ equals  $<\psi\mid\phi_{sj}><\phi_{sj}\mid H^{(0)} \mid\phi_{sj}><\phi_{sj}\mid\psi>/C_{s\_l}$ for nonzero $C_{s\_l}$ or zero if $C_{s\_l}$ is zero. The values of $a_{jl}^{(n-1)}$'s and $E_j^{(n-1)}$'s can be used to calculate $a_{jl}^{(n)}$'s and $E_j^{(n)}$'s. To achieve that, replacing $a_{jl}^{(n)}$ with $a_{jl}^{(n-1)}+da_{jl}^{(n-1)}$ and $E_j^{(n)}$ with $E_j^{(n-1)}+dE_j^{(n-1)}$, keeping the linear terms of $da_{jl}^{(n-1)}$, $dE_{jl}^{(n-1)}$, and $1/N$ but not the product terms, linear equations can be obtained from Eq.~\eqref{eq:9} and Eq.~\eqref{eq:10}. Some equations are long but the detailed expression is not necessary for the explanation of the analysis so are omitted here. 

The number of linear equations equals the number of unknown $da_{jl}^{(n-1)}$'s and $dE_{j}^{(n-1)}$'s. The inhomogeneous terms have magnitude on the order of $1/N$. If all the constraints are independent, then there will be a unique solution for $da_{jl}^{(n)}$'s and $dE_{j}^{(n)}$'s. In this situation, all the variables have magnitude on the order of $1/N$. If some of the constraints are dependent, then an extra constraint can be introduced which is to minimize a non-negative term $\sum\limits_{j,l=1}^{\infty}da_{jl}^{(n-1)*}\cdot da_{jl}^{(n-1)}$. With this extra constraint, a tentative solution is to set some number of variables to zero such that the rest can be uniquely solved. In this case, the magnitude of the solution is on the order of $1/N$ or higher. If this tentative solution does not minimize $\sum\limits_{j,l=1}^{\infty}da_{jl}^{(n-1)*}\cdot da_{jl}^{(n-1)}$, then a better solution exists. It follows this analysis that the optimum solution has magnitude on the order of $1/N$ or higher. That means the magnitude of the error terms is on the order of $1/N^2$ or higher. Thus the magnitude of the accumulated errors in $a_{jl}^{(n)}$ and $E_{j}^{(n)}$ is on the order of $1/N$ or higher so will vanish when $N$ approaches infinity. This proves the existence of exact solution for Eq.~\eqref{eq:2}. 
 
\subsection{Completeness and uniqueness of the solution}\label{subsec:2.2}

As indicated by Eq.~\eqref{eq:2} and Eq.~\eqref{eq:5}, the functions $\mid\psi_{sj}>$'s are linearly independent and their number equals the number of linearly independent $\mid\phi_{sl}>$'s. Since the latter forms a complete basis, the set of $\mid\psi_{sj}>$'s is also a complete basis. 	

The uniqueness of $\mid\psi_{sj}>$'s can be proven by contradiction. Assume there exists another set of $\mid\psi_{sj}'>$'s that also satisfies Eq.~\eqref{eq:2}, i.e., 
\begin{linenomath*}
\begin{equation} 
\begin{aligned}
<\psi^{'}_{sj}\mid\psi^{'}_{sk}>=\delta _{jk} , <\psi^{'}_j\mid H\mid\psi^{'}_k>=E^{'}_{j}\delta_ {jk}, \\
j,k\in \mathbb{N} \label{eq:11}
\end{aligned}
\end{equation} 
\end{linenomath*}
where $\mid\psi_{j}'>$ is the normalized $\mid\psi_{sj}'><\psi_{sj}'\mid\psi>$. Then an operator $U$ can be defined to map between the two normalized sets, i.e., 
\begin{linenomath*}
\begin{equation} 
\mid\psi^{'}_{j}> = U\mid\psi_{j}>, \;\;\; j\in \mathbb{N}. \label{eq:12}
\end{equation} 
\end{linenomath*}
Considering the separation of position variables of $s$ and $s\_$ in $\mid\psi_{j}>$ and $\mid\psi_{j}'>$, the operator $U$ can be written as the product of an operator $U_s$ for $s$ and an operator $U_{s\_}$ for $s\_$. The former maps $\mid\psi_{sj}>$ to $\mid\psi_{sj}'>$. The latter depends on the former and maps the normalized $<\psi_{sj}\mid\psi>$ to the normalized $<\psi_{sj}'\mid\psi>$. Due to the dependence, $U$ does not map a normalized $\mid\psi_{sj_1}><\psi_{sj_2}\mid\psi>$ to a normalized $\mid\psi_{sj_1}'><\psi_{sj_2}'\mid\psi>$ if $j_1$ does not equal $j_2$. This dependence requires $\mid\psi_{j}>$ to be a product of $\mid\psi_{sj}>$ and a function that takes $\mid\psi_{sj}>$ as an input to generate the normalized $<\psi_{sj}\mid\psi>$ when $U$ or $U^T$ is concerned. Similar requirement holds for $\mid\psi'_{j}>$. When $U$ operates on $\mid\psi_{j}>$, $\mid\psi_{sj}>$ of $\mid\psi_{j}>$ changes to $\mid\psi_{sj}'>$. This changes the input to the function in $\mid\psi_{j}>$ which will yield the normalized $<\psi_{sj}'\mid\psi>$. The combined change is that $\mid\psi_{j}>$ becomes $\mid\psi_{j}'>$, aligning with Eq.~\eqref{eq:12}. 

The normalized, orthogonal, and complete set of $\mid\psi_{sj}>$ and $\mid\psi_{sj}'>$ ensures an unitary $U_s$ so the adjoint operator $U_s^T$ maps $\mid\psi_{sj}'>$ back to $\mid\psi_{sj}>$. Further applying the definition of adjoint operator, the inner product $<\psi_{j}'\mid\psi_{j}'>$ or equivalently $<U\psi_{j}\mid\psi_{j}'>$ can be written as $<\psi_{j}\mid U^T\psi_{j}'>$. The adjoint operator $U^T$ equals $U_s^TU_{s\_}^T$. Because of the unitarity of $U_s^T$, $\mid\psi_{sj}'>$ of $\mid\psi_{j}'>$ is mapped to $\mid\psi_{sj}>$. This together with the function in $\mid\psi_{j}'>$ will change $\mid\psi_{j}'>$ to $\mid\psi_{j}>$, which requires $U_{s\_}^T$ to map the normalized  $<\psi_{sj}'\mid\psi>$ back to the normalized $<\psi_{sj}\mid\psi>$. Therefore $U^T$ is also an unitary operator. After the mapping, the inner product becomes $<\psi_{j}\mid\psi_{j}>$ which indeed equals the prior inner product $<\psi_{j}'\mid\psi_{j}'>$.

Now the second constraint of Eq.~\eqref{eq:11} becomes
\begin{linenomath*}
\begin{equation}
<\psi_j\mid U^THU\mid\psi_k>=E^{'}_{j}\delta_ {jk},\; j,k\in \mathbb{N} \label{eq:13}
\end{equation}
\end{linenomath*}
which also defines a self-adjoint operator $O'_s$ as a function of the sub-system. Since $O_s$ and $O'_s$ share the same eigenfunctions, they commute. Consequently $O'_s$ is a symmetry generator and there exists one-parameter unitary symmetry operator $\textnormal{exp}(i\epsilon O'_s)$ satisfying
\begin{linenomath*}
\begin{equation}
{\textnormal{exp}^T(i\epsilon O'_s)}O_s{\textnormal{exp}(i\epsilon O'_s)} = O_s \label{eq:14}
\end{equation}
\end{linenomath*}
where $\epsilon$ represents a real variable. The symmetry of $O_s$ originates from $H$, thus 
\begin{linenomath*}
\begin{equation}
{\textnormal{exp}^T(i\epsilon O'_s)}H{\textnormal{exp}(i\epsilon O'_s)} = H. \label{eq:15}
\end{equation}
\end{linenomath*}

Equation \eqref{eq:15} suggests that $H$ and $\textnormal{exp}(i\epsilon O'_s)$ have the same eigenfunctions. The eigenfunctions of the latter are in the $s$ domain. However, due to the interaction term $H_{int}$, the eigenfunctions of $H$ are in the combined $s$ and $s{\_}$ domain. To resolve this contradiction, $H$ and $\textnormal{exp}(i\epsilon O'_s)$ can not have the same eigenfunctions. That means $\textnormal{exp}(i\epsilon O'_s)$ does not exist, therefore the set of solution $\mid\psi_{sj}>$'s and $E_j$'s of Eq.~\eqref{eq:2} is unique. The uniqueness of $\mid\psi_{sj}>$'s and $E_j$'s guarantee the uniqueness of $O_s$.

\subsection{To calculate eigenvalues by perturbation method}\label{subsec:2.3}

For weak interaction $H_{int}$, the changes in the system eigenvalues $E_{j}$'s are small. The first order variations of $E_{j}$'s can be calculated by perturbation method. Part of the changes come from the changes in the eigenvalues $O_{sj}$'s of the sub-system. To quantify the changes of $O_{sj}$'s, rewrite $H_{int}$ as $\Delta W$ with $\Delta$ being a real value parameter, and denote $\mid\psi_{s\_j}>$ the normalized inner product $<\psi_{sj}\mid\psi>$ as well as $\mid\phi_{s\_j}>$ the normalized inner product $<\phi_{sj}\mid\psi>$. With these notations, the sub-system eigenvalue $O_{sj}$ can be expressed as
\begin{linenomath*}
\begin{equation}
O_{sj} = <\psi_{s\_j}\mid<\psi_{sj}\mid H_s+\Delta W\mid\psi_{sj}>\mid\psi_{s\_j}>. \label{eq:16}
\end{equation}
\end{linenomath*}
The wave function $\mid\psi_{sj}>$ and the eigenvalue $O_{sj}$ can also be expanded with respect to $\Delta$ as
\begin{linenomath*}
\begin{equation}
\mid\psi_{sj}> = \mid\phi_{sj}> + \Delta\mid\phi_{sj}^{(1)}> + \Delta^2\mid\phi_{sj}^{(2)}> + \cdot\cdot\cdot , \label{eq:17}
\end{equation}
\end{linenomath*}
and
\begin{linenomath*}
\begin{equation}
O_{sj} = E_{sj} + \Delta O_{sj}^{(1)} + \Delta^2 O_{sj}^{(2)} + \cdot\cdot\cdot . \label{eq:18}
\end{equation}
\end{linenomath*}

The right hand side of Eq.~\eqref{eq:17} can be used to calculate the norm of $\mid\psi_{sj}>$. Keeping the terms up to the first order of $\Delta$ and choosing the phase to make $<\phi_{sj}^{(1)}\mid\phi_{sj}>$ real, the normalization of $\mid\psi_{sj}>$ requires 
\begin{linenomath*}
\begin{equation}
<\phi_{sj}^{(1)}\mid\phi_{sj}> = 0 . \label{eq:19}
\end{equation}
\end{linenomath*}
Plugging Eq.(\ref{eq:17}, \ref{eq:18}) into Eq.~\eqref{eq:16} and making use of Eq.~\eqref{eq:1} and Eq.~\eqref{eq:19} yields 
\begin{linenomath*}
\begin{equation}
O_{sj}^{(1)} = <\phi_{s\_}\mid<\phi_{sj}\mid W \mid\phi_{sj}>\mid\phi_{s\_}>, \label{eq:20}
\end{equation}
\end{linenomath*}
so the increament of $O_{sj}$ due to $H_{int}$ equals 
\begin{linenomath*}
\begin{equation}
O_{sj} - E_{sj} = <\phi_{s\_}\mid<\phi_{sj}\mid H_{int} \mid\phi_{sj}>\mid\phi_{s\_}> \label{eq:21}
\end{equation}
\end{linenomath*}
for all $j\in \mathbb{N}$.

Corresponding to the expected value on the right hand side of Eq.~\eqref{eq:16}, the other component $E_{s\_j}$ of $E_{j}$ is 
\begin{linenomath*}
\begin{equation}
E_{s\_j} = <\psi_{s\_j}\mid<\psi_{sj}\mid H_{s\_} \mid\psi_{sj}>\mid\psi_{s\_j}> . \label{eq:22}
\end{equation}
\end{linenomath*}
Since $H_{s\_}$ does not act on $\mid\psi_{sj}>$, Eq.~\eqref{eq:22} reduces to  
\begin{linenomath*}
\begin{equation}
E_{s\_j} = <\psi_{s\_j}\mid H_{s\_} \mid\psi_{s\_j}>. \label{eq:23}
\end{equation}
\end{linenomath*}
To evaluate $E_{s\_j}$ in Eq.~\eqref{eq:23}, it is necessary to quantify the impact of $H_{int}$ to $\mid\psi_{s\_j}>$. This can be done by analyzing the solution of the Schr$\ddot{o}$dinger equation
\begin{linenomath*}
\begin{equation}
-i\hbar \frac{\partial}{\partial t}\mid\psi> = (H_s+\Delta W+H_{s\_})\mid\psi>. \label{eq:24}
\end{equation}
\end{linenomath*}

To do that, expand the wave function $\mid\psi>$ by the set of eigenfunctions $\mid\phi_{sk}>$'s as
\begin{linenomath*}
\begin{equation}
\mid\psi> = \sum\limits_{j}e^{i\omega_j t}\mid\phi_{sj}>\mid e^{-i\omega_j t}\psi_{s\_j}> \label{eq:25}
\end{equation}
\end{linenomath*}
where $\omega_j =E_{sj}/\hbar$ with $\hbar$ being the reduced Planck's constant. Making use of the first orthogonality condition in Eq.~\eqref{eq:2}, the Schr$\ddot{o}$dinger equation Eq.~\eqref{eq:25} can be written as
\begin{linenomath*}
\begin{equation}
-i\hbar \frac{\partial}{\partial t}\mid\tilde{\psi}_{s\_j}> = \sum\limits_{j'}{}c_k<\phi_{sj}\mid \Delta W \mid\phi_{sj'}>\mid\tilde{\psi}_{s\_j'}> + H_{s\_}\mid\tilde{\psi}_{s\_j}> \label{eq:26}
\end{equation}
\end{linenomath*}
where $c_j = e^{i(\omega_{j'}-\omega_j) t}$ and $\mid\tilde{\psi}_{s\_j}>= e^{-i\omega_j t}\mid\psi_{s\_j}>$. To the first order of $\Delta$, $<\psi_{s\_j}\mid<\phi_{sj}\mid \Delta W \mid\phi_{sj'}>\mid\psi_{s\_j'}> = \Delta O_{sj}^{(1)}\delta_{jj'}$. Making use of this approximation, Eq.~\eqref{eq:26} becomes
\begin{linenomath*}
\begin{equation}
<\tilde{\psi}_{s\_j}\mid-i\hbar\frac{\partial}{\partial t}\mid\tilde\psi_{s\_j}> = <\tilde{\psi}_{s\_j}\mid\Delta O_{sj}^{(1)}+H_{s\_}\mid\tilde{\psi}_{s\_j}> . \label{eq:27}
\end{equation}
\end{linenomath*}
Equation~\eqref{eq:27} indicates that to the first order, the impact of $H_{int}$ to $\mid\psi_{s\_j}>$ is to multiply the original wave component with $e^{i\Delta O_{sj}^{(1)}t/\hbar}$, i.e.,
\begin{linenomath*}
\begin{equation}
\mid\psi_{s\_j}> = \mid\phi_{s\_}>e^{i\Delta O_{sj}^{(1)}t/\hbar} . \label{eq:28}
\end{equation}
\end{linenomath*}

However, this multiplication does not change the expected value $H_{s\_}$. Therefore, to the first order of $\Delta$, the eigenvalue $E_{j}$ is
\begin{linenomath*}
\begin{equation}
E_{j} = <\phi_{s\_}\mid<\phi_{sj}\mid H_s + H_{int} + H_{s\_} \mid\phi_{sj}>\mid\phi_{s\_}> \label{eq:29}
\end{equation}
\end{linenomath*}
for all $j\in \mathbb{N}$.   

\section{Adjustment to hydrogen fine structure}\label{sec:3}

A hydrogen atom is a good object to test the correctness of equation \eqref{eq:2}. In the relativistic domain, the relative motion of the atom entangles with the center-of-mass motion. The impact of the latter to the energy level of the former can be calculated by Eq.~\eqref{eq:2} and Eq.~\eqref{eq:29}. The energy variation associated with $\mid\phi_{s\_}>$ can be also quantified. The overall impact has never been quantified before. It will result in adjustment to the fine structure of hydrogen atom. 

So far the successful and popular theory to quantify the fine structure is based on Dirac equation \cite{ 7}. Denote $m_e$ the electron mass, $\textbf{p}$ the momentum operator, $e$ the electron charge, $c$ the speed of light, $\epsilon_0$ the vacuum permittivity, $r_e$ the magnitude of the electron position $\textbf{r}_e$, $\textbf{S}$ the spin operator, $\textbf{L}$ the orbital angular momentum, and $\delta({\textbf{r})}$ Dirac's delta function. The Hamiltonian of an electron in a central potential $V$ can be derived as \cite{6} 
\begin{linenomath*}
\begin{equation}
H_e =\frac{\textbf{p}^2}{2m_e}+eV-\frac{\textbf{p}^4}{8m_e^3c^2}-\frac{e^2}{8\pi\epsilon_{0}m_e^2c^2r_e^3}\textbf{S}\cdot\textbf{L}-\frac{e^2\hbar^2}{8\pi\epsilon_{0}m_e^2c^2}\delta({\textbf{r}_e}) . \label{eq:30}
\end{equation}
\end{linenomath*}
When $V$ is the Coulomb potential between the electron and a nucleus at the origin, the first two terms of Eq.~\eqref{eq:30} together represents the commonly used Hamiltonian of the electron in a hydrogen atom in the non-relativistic domain. Denote $H^{(0)}$ the Hamiltonian. It defines a complete set of normalized eigenfunctions, i.e., 
\begin{linenomath*}
\begin{equation}
H^{(0)}\mid\phi_{nlm}> = E_{n}\mid\phi_{nlm}> \label{eq:31}
\end{equation}
\end{linenomath*}
where the energy eigenvalue $E_n$ equals $-\frac{m_ec^2\alpha^2}{2n^2}$ with $\alpha$ being the fine structure constant $\frac{e^2}{4\pi\epsilon_0\hbar c}$, the positive integer $n$ is the principal quantum number, the integer $l$ ranges from $0$ to $n-1$, and integer $m$ takes all the value between $-l$ and $l$. 

The rest three terms in Eq.~\eqref{eq:30} are much smaller than the first two so they can be treated as perturbation. The first order correction to the eigenvalues of $H$ can be obtained by using the perturbation method. For that purpose, the eigenfunctions $\mid\phi_{nlm}>$ and $\mid\phi_{nlm+1}>$ can be superimposed to be the common eigenfunctions of $\textbf{S}^2$, $\textbf{L}^2$, $\textbf{J}^2$, and $\textbf{J}_z$. Here $\textbf{J}$ represents the total angular momentum $\textbf{L}+\textbf{S}$ and $\textbf{J}_z$ is the $z$ component of $\textbf{J}$. The new functions are still eigenfunctions of $H^{(0)}$ with eigenvalues $E_{n}$'s. The value $j$ of the total angular momentum $\textbf{J}$ equals $\frac{1}{2}$ if $l$ equals zero and $l\pm\frac{1}{2}$ otherwise. The combined first order correction of the three terms is \cite{6}
\begin{linenomath*}
\begin{equation}
E_{n}^{(1)} = \frac{E_{n}\alpha^2}{n}(\frac{1}{j+\frac{1}{2}}-\frac{3}{4n}). \label{eq:32}
\end{equation}
\end{linenomath*}

\subsection{New relativistic Hamiltonian of hydrogen atom}\label{subsec:3.1}

In reality a proton nucleus is not confined at one location. The wave function of the electron entangles with that of the proton. The method to derive the Hamiltonian in Eq.~\eqref{eq:30} from Dirac equation assumed an independent wave function of the electron so does not hold anymore. To search for a new method, it is worthy noting that experiment results support predicted spectrum energy of relative motion, suggesting that the Hamiltonian of the relative motion should be derived instead. For that purpose, expand the relativistic kinetic energy $T_e$ of the electron as
 \begin{linenomath*}
\begin{equation}
\sqrt{\textbf{p}^2c^2+m_e^2c^4} = \frac{\textbf{p}^2}{2m_e} - \frac{\textbf{p}^4}{8m_e^3c^2} + \cdot\cdot\cdot. \label{eq:33}
\end{equation}
\end{linenomath*}
Denote $m_{p}$ the nucleus mass, $\textbf{r}_{p}$ the position variable of the nucleus, $\textbf{r}$ the relative position of the electron $\textbf{r}_{e}-\textbf{r}_{p}$ with respect to that of the nucleus, $m_t$ the total mass $m_e + m_p$, $\textbf{R}$ the position of the center-of-mass $(m_e{\textbf{r}_{e}}+m_p{\textbf{r}_{p}})/{m_t}$, and $T_{p}$ the kinetic energy operator $-\frac{{\hbar}^2}{2m_{p}}\triangledown^2_{\textbf{r}_{p}}$ of the nucleus. In the new coordinate system, $T_{p}$ changes to $- \frac{m_{p}{\hbar}^2}{2m^2_t}\triangledown^2_{\textbf{R}} -\frac{{\hbar}^2}{2m_{p}}\triangledown^2_{\textbf{r}} -\frac{1}{m_t}\textbf{p}_r\cdot\textbf{p}_R$ with $\textbf{p}_r$ equal to $-i\hbar\triangledown_{\textbf{r}}$ and $\textbf{p}_R$ equal to $-i\hbar\triangledown_{\textbf{R}}$. It is easy to verify that $\frac{\textbf{p}^2}{2m_e}+T_{p}$ equals $\frac{\textbf{p}_r^2}{2\mu}+\frac{\textbf{p}_R^2}{2m_t}$. 

The second term on the right hand side of Eq.~\eqref{eq:33} can be written as $- \frac{\textbf{p}_r^4}{8\mu^3c^2} +(\frac{\textbf{p}_r^4}{8\mu^3c^2} - \frac{\textbf{p}^4}{8m_e^3c^2})$. The summation $\frac{\textbf{p}_r^2}{2\mu} - \frac{\textbf{p}_r^4}{8\mu^3c^2}$ is approximately $\sqrt{\textbf{p}_r^2c^2+\mu^2c^4}$ with error on the order of $\frac{\textbf{p}^6}{16\mu^5c^4}$. So the atomic Hamiltonian $H$ can be approximated as the summation of $H_r$, which represents $\sqrt{\textbf{p}_r^2c^2+\mu^2c^4}+V$, $T_R$, and $\frac{\textbf{p}_r^4}{8\mu^3c^2} - \frac{\textbf{p}^4}{8m_e^3c^2}$. Since the third component of the summation is very small, the relative motion of the atom is nearly independent. By following Dirac's method \cite{7} to derive Dirac equation and the method \cite{6} to derive Eq.~\eqref{eq:30}, the relativistic form of the Hamiltonian of the relative motion can be obtained as
\begin{linenomath*}
\begin{equation}
H_r =\frac{\textbf{p}_r^2}{2\mu}+eV-\frac{\textbf{p}_r^4}{8\mu^3c^2}-\frac{e^2}{8\pi\epsilon_{0}\mu^2c^2r^3}\textbf{S}\cdot\textbf{L}_{\mu}-\frac{e^2\hbar^2}{8\pi\epsilon_{0}\mu^2c^2}\delta({\textbf{r}}) \label{eq:34}
\end{equation}
\end{linenomath*}
where $\textbf{L}_{\mu}$ is the angular momentum of the relative motion $\textbf{r}\times\textbf{p}_r$. The new hydrogen atom Hamiltonian $H$ is then
\begin{linenomath*}
\begin{equation}
H =\frac{\textbf{p}_r^2}{2\mu}+eV-\frac{\textbf{p}^4}{8m_e^3c^2}-\frac{e^2}{8\pi\epsilon_{0}\mu^2c^2r^3}\textbf{S}\cdot\textbf{L}_{\mu}-\frac{e^2\hbar^2}{8\pi\epsilon_{0}\mu^2c^2}\delta({\textbf{r}})+\frac{\textbf{p}_R^2}{2m_t}. \label{eq:35}
\end{equation}
\end{linenomath*}

The new Hamiltonian $H$ maintains the kinetic energy operator of the proton and the relativistic kinetic energy operator of the electron to the first order. In comparison, the original Hamiltonian $H_e$ omits the kinetic energy operator of the proton. The mass in the spin-orbital interaction term and the Darwin term of $H$ is the reduced mass while that in $H_e$ is the electron mass. Nevertheless, if an assumption is made for the proton to be infinitely massive, as was the case in deriving $H_e$, then the new Hamiltonian $H$ will be identical to the original $H_e$.   

\subsection{New energy level of fine structure}\label{subsec:3.2}

The first and second terms on the right hand side of Eq.~\eqref{eq:35} together is the well-known Hamiltonian of the relative motion of the hydrogen atom. It define a complete set of normalized eigenfunctions 
\begin{linenomath*}
\begin{equation}
(\frac{\textbf{p}_r^2}{2\mu}+eV)\mid\phi_{\mu,nlm}> = E_{\mu,n}\mid\phi_{\mu,nlm}> \label{eq:36}
\end{equation}
\end{linenomath*}
where the energy eigenvalue $E_{\mu,n}$ equals $-\frac{\mu c^2\alpha^2}{2n^2}$. The third to fifth terms on the right hand side of Eq.~\eqref{eq:35} affect the fine structure of hydrogen atoms. The third term can be expanded as
\begin{linenomath*}
\begin{equation}
-\frac{\textbf{p}^4}{8m_e^3c^2}=-\frac{1}{8m_e^3c^2}(\textbf{p}_r^4-4\frac{m_e}{m_t}\textbf{p}_r^3\cdot\textbf{p}_R+6\frac{m_e^2}{m_t^2}\textbf{p}_r^2\textbf{p}_R^2-4\frac{m_e^3}{m_t^3}\textbf{p}_r\cdot\textbf{p}_R^3+\frac{m_e^4}{m_t^4}\textbf{p}_R^4). \label{eq:37}
\end{equation}
\end{linenomath*}

Denote $\textbf{J}_{\mu}$ the total angular momentum $\textbf{L}_{\mu}+\textbf{S}$ and  $\textbf{J}_{\mu,z}$ its $z$ component. Similar to the calculation of Eq.~\eqref{eq:32}, the eigenfunctions $\mid\phi_{\mu,nlm}>$ and $\mid\phi_{\mu,nlm+1}>$ can be superimposed to be the common eigenfunctions of $\textbf{S}^2$, $\textbf{L}_{\mu}^2$, $\textbf{J}_{\mu}^2$, and $\textbf{J}_{\mu,z}$. The combined energy shift of the fine structure due to $-\frac{1}{8m_e^3c^2}\textbf{p}_r^4$, $-\frac{e^2}{8\pi\epsilon_{0}\mu^2c^2r^3}\textbf{S}\cdot\textbf{L}_{\mu}$, and $-\frac{e^2\hbar^2}{8\pi\epsilon_{0}\mu^2c^2}\delta({\textbf{r}})$ can be calculated as
\begin{linenomath*}
\begin{equation}
E_{n}^{'(1)} = \frac{E_{\mu,n}\alpha^2}{n}(\frac{1}{j+\frac{1}{2}}-\frac{3}{4n})-\frac{E_{\mu,n}\alpha^2}{n}(\frac{1}{j}-\frac{3}{4n})(1-\frac{m_p^3}{m_t^3}). \label{eq:38}
\end{equation}
\end{linenomath*}

However, the second to the fourth terms on the right hand side of Eq.~\eqref{eq:37} contain operators of $\textbf{r}$ and $\textbf{R}$. They represent the interaction between the relative motion and the center-of-mass motion. Their impact to the energy level of the fine structure needs to be quantified too. Applying Eq.~\eqref{eq:2} to the sub-system of the relative motion of the hydrogen atom, the first order impact equals the inner product of $H_{int}$ on the right hand side of Eq.~\eqref{eq:29}. The parity of the wave function $\mid\phi_{sj}>$, which is the linear combination of $\mid\phi_{\mu,nlm}>$ and $\mid\phi_{\mu,nlm+1}>$, depends on $l$. When $\textbf{r}$ changes sign, i.e, $\textbf{r}\rightarrow -\textbf{r}$, the linear momentum $\textbf{p}_r$ and its cubic $\textbf{p}_r^3$ change sign, and $\mid\phi_{sj}>$ becomes $(-1)^l\mid\phi_{sj}>$. So the inner products of $\textbf{p}_r$ and $\textbf{p}_r^3$ are integral of odd functions. The results are zero. The inner product of the remaining term $-\frac{3\textbf{p}_r^2\textbf{p}_R^2}{4m_em_t^2c^2}$ can be calculated as $\frac{3m_pE_{\mu,n}E_{cm}}{m_t^2c^2}$ with $E_{cm}$ being the kinetic energy of the center-of-mass motion. 

The energy of absorbed or emitted photon associated with the transition between $\mid\psi_{n}>$ and $\mid\psi_{k}>$ equals the energy difference between $E_{\mu,n}+E_n^{'(1)}+\frac{3m_pE_{\mu,n}E_{cm}}{m_t^2c^2}$ and $E_{\mu,k}+E_k^{'(1)}+\frac{3m_pE_{\mu,k}E_{cm}}{m_t^2c^2}$, meaning that the total energy adjustment due to the fine structure is
\begin{linenomath*}
\begin{equation}
E_{n}^{''(1)} = \frac{E_{\mu,n}\alpha^2}{n}(\frac{1}{j+\frac{1}{2}}-\frac{3}{4n})-\frac{E_{\mu,n}\alpha^2}{n}(\frac{1}{j}-\frac{3}{4n})(1-\frac{m_p^3}{m_t^3})+\frac{3m_pE_{\mu,n}E_c}{m_t^2c^2}. \label{eq:39}
\end{equation}
\end{linenomath*}
The impact of the center-of-mass motion to the energy level of the fine structure is quite small. For example, for $100eV$ kinetic energy $E_{cm}$, the ratio $\frac{3m_pE_{cm}}{m_t^2c^2}$ is only $3.2\times10^{-7}$. 

The difference between Eq.~\eqref{eq:39} and Eq.~\eqref{eq:32} can be illustrated by using an example. Consider the fine structure of $2P_{1/2}-1S_{1/2}$ transition with $100eV$ kinetic energy $E_{cm}$. The fine structure energy adjustment from Eq.~\eqref{eq:32} is $1.24\times10^{-4}eV$. In comparison, that from Eq.~\eqref{eq:39} is $1.26\times10^{-4}eV$. The difference is $1.95\times10^{-6}eV$. The last term on the right hand side of Eq.~\eqref{eq:39}, which represents the impact of the center-of-mass motion, contributes $3.3\times10^{-6}eV$. 

\section{To quantify position measurement results by photon emission and absorption}\label{sec:4}

Position measurements are important experiments in quantum mechanics. The interference patterns of accumulated position measurements have been successfully observed for electrons \cite{8,9}, neutrons \cite{10}, and neon atoms \cite{11}. These experiments have convincingly demonstrate the existence of matter waves and particle wave duality. The interference patterns can be described as wave propagation subject to boundary conditions of the diffractors \cite{12,13}. In quantum mechanics, the detection of one location of one incident electron is explained as the result of wave function collapsing to a position eigenfunction, and the frequency to get the same measurement result is quantified by Born's rule. The explanation is acceptable for empirical purpose.    

From the requirements of a consistent theory, that explanation, however, is questionable. First, the kinetic energy of any position eigenfunction is infinite so a wave function can not collapse to such eigenfunction. A feasible eigenfunction can only be a quasi-position eigenfunction. Whatever mechanism resulting in that should guarantee the uniqueness of the set of eigenfunctions. Second, the explanation omits the physical property of the detector and fails to account for other aspects of the measurement results such as the moment when a collapse process finishes, the number of emitted photons, the distribution of the emitted photons in space, etc.   

To overcome the weakness of the empirical explanation, the energy eigenvalue equation will be applied to analyze position measurements. Because the paper \cite{8} provides some necessary details of measurement setup and measurement results, analysis will be carried out to quantify the results reported in that paper.  

In that experiment, each of the incident electron had $50Kev$ kinetic energy with about $1um$ length of the wave packet. The electron struck a fluorescent film at one time and on average produced about $500$ photons. The fluorescent material was not specified. By considering the time when the experiment was carried out, there is high chance for the material to be doped zinc sulfide (ZnS). 

Studies, e.g., \cite{14} have shown that after an electron enters a solid, it will be scattered by the atoms in the solid \cite{14}. Inelastic scattering will cause the incident electron to lose kinetic energy. In terms of waveform, this phenomena can be described as the slowing down of the group velocity of the electron wave packet during its course of traveling into the detector. For any region of the detector, a sub-system can be defined that describes the relative motion of the incident electron and the atoms in the region. The reference coordinate can be that of some nucleus in the sub-system. The detector being a macroscopic object means that the location of the nucleus is well defined. For convenience, use that location as the origin of the coordinate system so the coordinates of other particles in the sub-system are the coordinates of the relative motion. According to the proof in Section \ref{sec:2}, the system Hamiltonian and wave function determines a unique set of eigenfunctions for the sub-system. The system wave function can be expanded by the set of eigenfunctions. For an incident electron with high kinetic energy, the expansion coefficients of bound eigenfunctions may be extremely small. When the electron is close to its maximum penetration depth, however, the electrons in the wavefront has lost enough kinetic energy so the magnitude of the expansion coefficients of bound eigenfunctions significantly increase. 

Generally the interaction between a charge in any sub-system and electromagnetic field will cause spontaneous emission. Denote $\mid\psi_{si_b}>$ an initial bound eigenfunction, $\mid\psi_{sf_b}>$ the final bound eigenfunction, $E_{si}$ the energy of the initial state, $E_{sf}$ the energy of the final state, $\hat{H}_{int}$ the interaction Hamiltonian between the sub-system and photons, $\hat{\textbf{A}}$ the vector potential operator of the photon, $\hat{\textbf{p}}$ the momentum operator of the incident electron, $\hat{\textbf{r}}$ the position vector of the photon wave, $\textbf{k}_{if}$ the wave vector of the photon, $\textbf{d}$ the transition dipole moment $<\psi_{sf_b}\mid\hat{\textbf{r}}\mid\psi_{si_b}>$, $\omega_{if}$ the angular frequency of a photon with energy equal to the energy difference between the initial and final eigenfunctions, and $\rho(E)$ the density of the final state per unit energy. According to $\textit{Fermi's Golden rule}$ \cite{15}, the transition rate from $\mid\psi_{si_b}>$ to $\mid\psi_{sf_b}>$ equals
\begin{linenomath*}
\begin{equation}
\kappa = \frac{2\pi}{\hbar}\mid<\psi_{sf_b}\mid\hat{H}_{int}\mid\psi_{si_b}>\mid^2\rho(E). \label{eq:40}
\end{equation}
\end{linenomath*}
   
The interaction term $\hat{H}_{int}$ equals $\frac{e}{2m_e}(\hat{\textbf{A}}\cdot\hat{\textbf{p}}+\hat{\textbf{p}}\cdot\hat{\textbf{A}})$. In free space, a single mode vector potential operator of a photon in direction $\textbf{n}_{\textbf{k}\sigma}$ with polarization $\sigma$ can be written as $\sqrt{\hbar/(2\epsilon_0\omega)}(\hat{a}_{\textbf{k}\sigma}e^{i(\textbf{k}\cdot\hat{\textbf{r}}-\omega t)}+\hat{a}_{\textbf{k}\sigma}^{\dag}e^{-i(\textbf{k}\cdot\hat{\textbf{r}}-\omega t)})\textbf{n}_{\textbf{k}\sigma}$. Here $\hat{a}_{\textbf{k}\sigma}$ and $\hat{a}_{\textbf{k}\sigma}^{\dag}$ represent annihilation and creation operator, respectively. The photon density of state $\rho(E)$ equals $8\pi{\hbar}^2\omega_{if}^2/(hc)^3$. Thus, the transition element $<\psi_{sf_b}\mid\hat{H}_{int}\mid\psi_{si_b}>$ becomes $\sqrt{\hbar/(2\epsilon_0\omega_{if})}\textbf{n}\cdot\left(2<\psi_{sf_b}\mid e^{i\textbf{k}_{if}\cdot\hat{\textbf{r}}}\hat{\textbf{p}}\mid\psi_{si_b}>+<\psi_{sf_b}\mid\hat{\textbf{p}}e^{i\textbf{k}_{if}\cdot\hat{\textbf{r}}}\mid\psi_{si_b}>\right)$ with $\textbf{n}$ being a unit vector in space. In case $e^{i\textbf{k}_{ij}\cdot\hat{\textbf{r}}}$ is approximately constant in the effective domain of $\mid\psi_{si_b}>$ and $\mid\psi_{sf_b}>$, then the second term in the parenthesis can be ignored, and the first term approximates $-2im_e\omega_{if}\textbf{d}$. In this situation, the transition rate from $\mid\psi_{si_b}>$ to $\mid\psi_{sf_b}>$ due to spontaneous photon emission simplifies to the following equation \cite{6}.  
\begin{linenomath*}
\begin{equation}
\kappa = \frac{4\alpha\omega_{if}^3\overline{{\mid \textbf{d}\mid}^2}}{3c^2}. \label{eq:41}
\end{equation}
\end{linenomath*}

This analysis provides a good foundation for quantifying the transition rates due to spontaneous photon emission. The transition in the position measurement, however, is more complicated. First of all, Eq.~\eqref{eq:41} was derived by using plane wave function in the vector potential operator $\hat{\textbf{A}}$. However, finite lifetime of the initial state broadens the spectral line width \cite{6}, causing Lorentzian shape frequency distribution in the spectral line \cite{6}. So a vector potential should not be assumed monochromatic. The Lorentzian frequency distribution and the factor $e^{i\textbf{k}\cdot\hat{\textbf{r}}}$ together will cause vector potential to decay exponentially in space. Also the finite distribution in space will determine the effective domain of the sub-system. These new features and their impact to the transition rate needs to be fully analyzed.

In addition to that, the initial wave function in the new transition is the wave function $\mid\psi>$ of the system instead of a bound eigenfunction $\mid\psi_{si_b}>$ of the sub-system, and the final wave function is $\mid\psi_{sf_b}>\mid\psi_{s\_f_b}>$ instead of $\mid\psi_{sf_b}>$. It is necessary to analyze whether $\textit{Fermi's Golden rule}$ is applicable to this case. 

According to the new theory, the transition rate directly affects the photon size and the number of emitted photons in the position measurement. The photon size is also the effective size of the sub-system. The photons are quickly absorbed by valence electrons and cause those electrons to transition to the conduction band. The excited electrons in the conduction band eventually recombine with holes, yielding position measurement results. The relevant analysis, derivation, and numerical results will be provided in the rest sub-sections.

\subsection{Photon localization }\label{subsec:4.1}
 
The lifetime $\tau$ of a state is the inverse of the transition rate from that state to all other states. A finite lifetime causes multiple frequency components in the spectral line. The frequency distribution can be characterized by a Lorentzian function $L(\omega)$ with frequency dependent factor as $\frac{1}{(\omega-\omega_{if})^2+\frac{1}{4\tau^2}}$ \cite{6}. Due to the photon dispersion relation $\omega = c\mid\textbf{k}\mid$ between the angular frequency $\omega$ and the magnitude $\mid\textbf{k}\mid$ of the wave vector $\textbf{k}$, there will be an equivalent Lorentzian function $L^{'}(\textbf{k})$. 

The curl of a vector potential $\textbf{A}$ is the magnetic field and its time derivative is part of the electric field. To have Lorentzian spectrum lineshape, $\textbf{A}$ should have the distribution of $L^{'}(\textbf{k})$. However, $L^{'}(\textbf{k})$ is not simply $L(\omega)$ with replacement $\omega = c\mid\textbf{k}\mid$. The number of plane waves between $\omega$ and $\omega+d\omega$ is proportional to $L(\omega)d\omega$. If the summation is done in terms of the momentum, the momentum distribution will be $\frac{1}{(2\pi)^3}L^{'}(\textbf{k})d^3\textbf{k}$. For this to be the same as $L(\omega)d\omega$, $L^{'}(\textbf{k})$ needs to differ from $L(\omega)$. After carefully analysis, an approximate $L^{'}(\textbf{k})$ with high accuracy is found to be
\begin{linenomath*}
\begin{equation}
L^{'}(\textbf{k}) \approx  \frac{16\pi c\tau\mid\textbf{k}_{if}\mid}{\mid\textbf{k}\mid\left((\textbf{k}-\textbf{k}_{if})^2+\frac{1}{4c^2\tau^2}\right)^2}  . \label{eq:42}
\end{equation}
\end{linenomath*}
The approximation in Eq.~\eqref{eq:42} is due to the second term in $\frac{1}{(\mid\textbf{k}\mid-\mid\textbf{k}_{if}\mid)^2+\frac{1}{4c^2\tau^2}}+\frac{1}{(\mid\textbf{k}\mid+\mid\textbf{k}_{if}\mid)^2+\frac{1}{4c^2\tau^2}}$ being dropped. This approximation is valid as long as $\mid\textbf{k}_{if}\mid$ is much larger than $1/(2c\tau)$. Usually this is satisfied since the half width of spectral line is much smaller than the photon frequency. It can be verified that $L^{'}(\textbf{k})$ is normalized, i.e., 
\begin{linenomath*}
\begin{equation}
\frac{1}{8\pi ^3}\iiint L^{'}(\textbf{k}) \, d^3\textbf{k}=1 . \label{eq:43}
\end{equation}
\end{linenomath*}

Adding up all the plane waves $e^{i\textbf{k}\cdot\hat{\textbf{r}}}$ with different weight $L^{'}(\textbf{k})$ is equivalent to calculating the three dimensional inverse Fourier transform of $L^{'}(\textbf{k})$, i.e., the transform result $\textbf{A}_{packet}(\hat{\textbf{r}})$ is  
\begin{linenomath*}
\begin{equation}
\textbf{A}_{packet}(\hat{\textbf{r}}) = \frac{C\textbf{n}}{8\pi^3}\iiint \frac{16\pi c\tau\mid\textbf{k}_{if}\mid e^{i\textbf{k}\cdot\hat{\textbf{r}}}}{\mid\textbf{k}\mid\left(\mid\textbf{k}-\textbf{k}_{if}\mid^2+\frac{1}{4c^2\tau^2}\right)^2} \, d^3\textbf{k}  \label{eq:44}
\end{equation}
\end{linenomath*}
where the coefficient $C$ is to be determined. This integration can be evaluated by using spherical coordinates. If an approximation is made to equate $\mid\textbf{k}_{if}\mid$ in the numerator and $\mid\textbf{k}\mid$ on the denominator of the integrand, then a closed form solution can be obtained as 
\begin{linenomath*}
\begin{equation}
\textbf{A}_{packet}(\hat{\textbf{r}}) \approx Ce^{i\textbf{k}_{if}\cdot(\hat{\textbf{r}}-\hat{\textbf{r}}_o)-\frac{\mid\hat{\textbf{r}}-\hat{\textbf{r}}_o\mid}{2c\tau} }\textbf{n}  \label{eq:45}
\end{equation}
\end{linenomath*}
with $\hat{\textbf{r}}_o$ representing the center location of the wave packet potential. As can be seen, the wave vector $\textbf{A}_{packet}(\hat{\textbf{r}})$ is a plane wave $e^{i\textbf{k}_{if}\cdot(\hat{\textbf{r}}-\hat{\textbf{r}}_o)}\textbf{n}$ modulated by an amplitude $Ce^{-\frac{\mid\hat{\textbf{r}}-\hat{\textbf{r}}_o\mid}{2c\tau} }$ that decays exponentially at location away from the center. This decay rate is the maximum rate that is allowed by the theory \cite{16}. The decay localizes the wave vector. The field essentially exists in a sphere with radius $2n_kc\tau$ that is $n_k$ times larger than $2c\tau$. For example, if $n_k$ equals 5, the magnitude of the wave vector on the sphere surface is less than $1\%$ of the magnitude at the center. 

To determine the coefficient $C$, it is required that the atomic spontaneous emission rate due to localized vector potential $\textbf{A}_{packet}(\hat{\textbf{r}})$ should be the same as that due to the plane wave vector potential. Adopting the same reasoning as in the latter case, the term $\mid<\psi_{sf_b}\mid\hat{H}_{int}\mid\psi_{si_b}>\mid^2$ with $\hat{H}_{int}$ being $\frac{e}{2m_e}(2\textbf{A}_{packet}(\hat{\textbf{r}})\cdot\hat{\textbf{p}}+\hat{\textbf{p}}\cdot\textbf{A}_{packet}(\hat{\textbf{r}}))$ is found to be $\frac{e^2}{m_e^2V_{\textbf{A}}}\mid<\psi_{sf_b}\mid\textbf{n}\cdot\hat{\textbf{p}}\mid\psi_{si_b}>\mid^2\iiint \limits_{V_{\textbf{A}}} \mid\textbf{A}_{packet}(\hat{\textbf{r}})\mid^2 \, d^3\hat{\textbf{r}}$. The volume $V_{\textbf{A}}$ of the effective domain is $\frac{4\pi}{3}(2n_kc\tau)^3$.  The integral $\iiint \limits_{V_{\textbf{A}}} \mid\textbf{A}_{packet}(\hat{\textbf{r}})\mid^2 \, d^3\hat{\textbf{r}}$ can be easily evaluated as $\pi (2c\tau)^3(1-\frac{1}{2}((2n_k+1)^2+1)e^{-2n_k})$. When $n_k$ is relative large, e.g., $n_k\geq 5$, the term is small enough to be discarded. So for $\frac{e^2}{m_e^2V_{\textbf{A}}}\mid<\psi_{sf_b}\mid\textbf{n}\cdot\hat{\textbf{p}}\mid\psi_{si_b}>\mid^2\iiint \limits_{V_{\textbf{A}}} \mid\textbf{A}_{packet}(\hat{\textbf{r}})\mid^2 \, d^3\hat{\textbf{r}}$ to equal $\frac{\hbar}{2\epsilon_0\omega_{if}}\frac{e^2}{m_e^2}\mid<\psi_{sf_b}\mid\textbf{n}\cdot\hat{\textbf{p}}\mid\psi_{si_b}>\mid^2$, $C^2$ can only be
\begin{linenomath*}
\begin{equation}
C^2 = \frac{2\hbar}{3\epsilon_0\omega_{if}}n_k^3 . \label{eq:46}
\end{equation}
\end{linenomath*} 
Under this condition, the transition rate due to the localized vector potential $\textbf{A}_{packet}(\hat{\textbf{r}})$ remains the same.  

After the photon is generated, the center of the wave packet moves in the direction of $\textbf{k}_{if}$ with speed of light. To account for the time dependence, the center $\hat{\textbf{r}}_o$ should be replaced with $\hat{\textbf{r}}_o(t)$ that equals $\hat{\textbf{r}}_o-\frac{\textbf{k}_{if}}{\mid\textbf{k}_{if}\mid}ct$. Thus the complete form of wave packet $\textbf{A}_{packet}(\hat{\textbf{r}},t)$ in free space becomes
\begin{linenomath*}
\begin{equation}
\textbf{A}_{packet}(\hat{\textbf{r}},t) \approx \sqrt{\frac{2\hbar}{3\epsilon_0\omega_{if}}n_k^3}e^{i\textbf{k}_{if}\cdot(\hat{\textbf{r}}-\hat{\textbf{r}}_o(t))-\frac{\mid\hat{\textbf{r}}-\hat{\textbf{r}}_o(t)\mid}{2c\tau} }\textbf{n} . \label{eq:47}
\end{equation}
\end{linenomath*}

\subsection{Transition rate of the incident electron}\label{subsec:4.2}

At this moment, the constituent of the sub-system is unknown except that it contains the incident electron and some atoms in the detector. Considering the spherical symmetry of the vector potential, the sub-system is assumed to be a sphere with center coincident with that of the vector potential. Denote $V_s$ the volume of the sub-system, which can be written as $\frac{4\pi}{3}(2b_kc\tau)^3$ with $2b_kc\tau$ being the radius of the sub-system. Although the volume $V_s$ is finite, the lattice maintains the lattice periodicity of the bulk material except at the boundary. The impact of the surroundings is mainly on the particles at the boundary, and the preferred basis of the sub-system is largely determined by the particles of the sub-system. For a sphere with radius on the order of $um$, the number of contained primitive cells is significant. The large amount and the periodicity define the sub-system's own valence band and conduction band, which differ from those of the bulk material but resemble them. In the expansion of the wave function $\mid\psi>$, generally there will be nonzero components of those valence band and conduction band eigenfunctions.

The energy of the incident electron in the wavefront near the maximum penetration depth falls in the energy range of the conduction band first. So its components of valence states can be ignored and the likely transition can only start from conduction states. However, transitions from conduction states to valence states are greatly limited by the low density of hole states in the valence band. The remedy is to create holes to enable fast transitions. On one hand, the photon-charge interaction causes transition to the valence band. On the other hand, the interaction between those photons and  valence band electrons elevate some of them to the conduction band, leaving behind holes. The latter can be a very fast process. The two processes together resolve the issue of low hole state density.   

When the final state $\mid\psi_{sf_b}>$ is an eigenstate of the incident electron in the valence band of the sub-system, for any component of $\mid\psi>$ that contains a unbound state of the incident electron, since it has negligible overlap with $\mid\psi_{sf_b}>$, its transition to $\mid\psi_{sf_b}>$ can be ignored. In other words, only the components of bound conduction states of the incident electron need to be considered. 

For any bound state $\mid\psi_{si_b}>$ in the conduction band, Bloch's theorem \cite{16} can be used to analyze its transition to $\mid\psi_{sf_b}>$. According to Bloch's theorem, the eigenfunction $\mid\psi_{si_b}>$ can be written as the product of $e^{i\textbf{k}_{i_b}\cdot({\hat{\textbf{r}}-\hat{\textbf{r}}_o})}$ and a periodic function $\mid u_{si_b}>$ with periodicity of the lattice. Similarly, $\mid\psi_{sf_b}>$ can be written as $e^{i\textbf{k}_{f_b}\cdot({\hat{\textbf{r}}-\hat{\textbf{r}}_o})}\mid u_{sf_b}>$ where $\mid u_{sf_b}>$ is also a periodic function with periodicity of the lattice. Applying this property to the element $<\psi_{sf_b}\mid\hat{H}_{int}\mid\psi_{si_b}>$ yields $\frac{e}{2m_e}\left(2<\psi_{sf_b}\mid\textbf{A}_{packet}(\hat{\textbf{r}})\cdot\hat{\textbf{p}}\mid\psi_{si_b}>+<\psi_{sf_b}\mid\hat{\textbf{p}}\cdot\textbf{A}_{packet}(\hat{\textbf{r}})\mid\psi_{si_b}>\right)$. Denote $\textbf{A}_{sum}$ the summation $\sum_{\hat{\textbf{r}}-\hat{\textbf{r}}_o=\textbf{a}_j}e^{i(-\textbf{k}_{f_b}+\textbf{k}_{i_b})\cdot(\hat{\textbf{r}}-\hat{\textbf{r}}_o)}\textbf{A}_{packet}$ at the locations $\textbf{a}_j$'s of all the primitive cells in the sub-system. The first term becomes $\frac{e}{m_e}<u_{sf_b}\mid\hbar\textbf{k}_{f_b}\cdot\textbf{A}_{sum}(\hat{\textbf{r}})\mid u_{si_b}>$ plus $\frac{e}{m_e}<u_{sf_b}\mid\textbf{A}_{sum}\cdot\hat{\textbf{p}}\mid u_{si_b}>$. Equation~\eqref{eq:44} can be used to evaluate $\hat{\textbf{p}}\cdot\textbf{A}_{packet}$ in the second term. The result is the same as the right hand side of Eq.~\eqref{eq:44} except a new multiplier $\textbf{k}$ to the integrand. Similar to the approximation in evaluating the integral in Eq.~\eqref{eq:44}, this new multiplier can be approximated as $\textbf{k}_{if}$, so $\textbf{p}\cdot\textbf{A}_{packet}$ approximates $\textbf{k}_{if}\cdot\textbf{A}_{packet}$. Then the second term can be written as $\frac{e}{2m_e}<u_{sf_b}\mid\hbar\textbf{k}_{if}\cdot\textbf{A}_{sum}\mid u_{si_b}>$. 

In the cases where $(-\textbf{k}_{f_b}+\textbf{k}_{i_b}+\textbf{k}_{if})\cdot(\hat{\textbf{r}}-\hat{\textbf{r}}_o)$ varies slowly within all primitive cells, the summation $\sum_{\hat{\textbf{r}}-\hat{\textbf{r}}_o=\textbf{a}_j}e^{i(-\textbf{k}_{f_b}+\textbf{k}_{i_b}\cdot(\hat{\textbf{r}}-\hat{\textbf{r}}_o)}\textbf{A}_{packet}(\hat{\textbf{r}})$ becomes an integral $I_{\textbf{A}}$ or $\frac{1}{V_{cell}}\iiint \limits_{V_s} e^{i(-\textbf{k}_{f_b}+\textbf{k}_{i_b})\cdot\hat{\textbf{r}}}\textbf{A}_{packet}(\hat{\textbf{r}}) \, d^3\hat{\textbf{r}}$ with $V_{cell}$ being the volume of a unit cell. Substituting $\textbf{A}_{packet}$ with its expression, the integral $I_{\textbf{A}}$ becomes 
\begin{linenomath*}
\begin{equation}
 I_{\textbf{A}} = \frac{C\textbf{n}}{V_{cell}}\iiint \limits_{V_s} e^{i(-\textbf{k}_{f_b}+\textbf{k}_{i_b}+\textbf{k}_{if})\cdot\hat{\textbf{r}}-\frac{\mid\hat{\textbf{r}}\mid}{2c\tau}} \, d^3\hat{\textbf{r}} . \label{eq:48}
\end{equation}
\end{linenomath*}
The above integration can be calculated analytically. When $e^{-b_k}$ is much less than one, replacing the finite domain with an infinite domain introduces small errors. Keeping most of the error terms, a simple and accurate expression of the integration is 
\begin{linenomath*}
\begin{equation}
I_{\textbf{A}} \approx 8\pi\frac{C\textbf{n}}{V_{cell}}\frac{\frac{1}{2c\tau}}{(k'^2+\frac{1}{4c^2\tau^2})^2}(1-\gamma)  \label{eq:49}
\end{equation}
\end{linenomath*}
where $k'$ equals $\mid -\textbf{k}_{f_b}+\textbf{k}_{i_b}+\textbf{k}_{if}\mid$ and $\gamma = \frac{(b_k+1)^2+1}{2}e^{-b_k}$. Clearly $\mid I_{\textbf{A}}\mid$ reduces quickly when $k'$ increases. For example, if $k'$ equals $2c\tau$, then $\mid I_{\textbf{A}}\mid$ drops to $4\%$ of its value at $k'=0$. 

This property selects transitions between eigenfunctions. The primary difference between $\mid\psi_{s\_i_b}>$ and $\mid\psi_{s\_f_b}>$ is the unbound states of the incident electron in $\mid\psi_{s\_i_b}>$. As pointed out earlier, their transitions to the bound state $\mid\psi_{sf_b}>$ are negligible. So the transition rate between normalized $\mid\psi_{si_b}>\mid\psi_{s\_i_b}>$ and $\mid\psi_{sf_b}>\mid\psi_{s\_f_b}>$ is essentially that between $\mid\psi_{si_b}>$ and $\mid\psi_{sf_b}>$. As long as the energy of the photons released by the incident electron matches the energy difference of the two eigenstates $\mid\psi_{si_b}>$ and $\mid\psi_{sf_b}>$, $\textit{Fermi's Golden rule}$ can be applied to quantify the transition rate.

On the other hand, $\mid\textbf{k}_{i_b}\mid$ and $\mid\textbf{k}_{f_b}\mid$ are usually much larger than $\mid\textbf{k}_{if}\mid$ and $1/(2c\tau)$. That means $\textbf{k}_{f_b}\approx\textbf{k}_{i_b}$. It follows that $<u_{sf_b}\mid\hbar\textbf{k}_{f_b}\cdot\sum_{\hat{\textbf{r}}-\hat{\textbf{r}}_o=\textbf{a}_j}e^{i(-\textbf{k}_{f_b}+\textbf{k}_{i_b})\cdot(\hat{\textbf{r}}-\hat{\textbf{r}}_o)}\textbf{A}_{packet}(\hat{\textbf{r}})\mid u_{si_b}>$ approximately equals $\hbar\textbf{k}_{f_b}\cdot I_{\textbf{A}}<\psi_{sf_b}\mid\psi_{si_b}>$, which is zero due to the orthogonality between $\mid\psi_{si_b}>$ and $\mid\psi_{sf_b}>$. So both $\frac{e}{m_e}<u_{sf_b}\mid\hbar\textbf{k}_{f_b}\cdot\textbf{A}_{sum}\mid u_{si_b}>$ and $\frac{e}{2m_e}<u_{sf_b}\mid\hbar\textbf{k}_{if}\cdot\textbf{A}_{sum}\mid u_{si_b}>$ can be ignored. 

The only remaining component $\frac{e}{m_e}<u_{sf_b}\mid\textbf{A}_{sum}\cdot\hat{\textbf{p}}\mid u_{si_b}>$ can be written as $\frac{e\mid I_{\textbf{A}}\mid}{m_e}<u_{sf_b}\mid\textbf{n}\cdot\hat{\textbf{p}}\mid u_{si_b}>$ or $\frac{e\mid I_{\textbf{A}}\mid V_{cell}}{m_eV_s}<\sqrt{\frac{V_s}{V_{cell}}}u_{sf_b}\mid\textbf{n}\cdot\hat{\textbf{p}}\mid\sqrt{\frac{V_s}{V_{cell}}} u_{si_b}>$. Using the definition of $E_p$ \cite{17}, $\mid<\sqrt{\frac{V_s}{V_{cell}}}u_{sf_b}\mid\textbf{n}\cdot\hat{\textbf{p}}\mid\sqrt{\frac{V_s}{V_{cell}}} u_{si_b}>\mid^2$ can be replaced with $m_eE_p/6$. Thus the transition rate $\kappa_{if}$ from $\mid\psi_{si_b}>$ to $\mid\psi_{sf_b}>$ is   
\begin{linenomath*}
\begin{equation}
\kappa_{if} = \frac{2\pi}{\hbar}\frac{e^2}{m_e^2}\frac{2\hbar}{3\epsilon_0\omega_{if}}\frac{64\pi^2n_k^3}{V_s^2}\frac{\frac{1}{4c^2\tau^2}(1-\gamma)^2}{(k'^2+\frac{1}{4c^2\tau^2})^4} \frac{8\pi{\hbar}^2{\omega_{if}}^2}{h^3c^3} , \label{eq:50}
\end{equation}
\end{linenomath*}
which can be simplified to
\begin{linenomath*}
\begin{equation}
\kappa_{if} = \frac{4\alpha\omega_{if}}{c^2}\frac{E_p}{6m_e}\frac{48n_k^3}{(2c\tau)^8b_k^6}\frac{(1-\gamma)^2}{(k'^2+\frac{1}{4c^2\tau^2})^4} . \label{eq:51}
\end{equation}
\end{linenomath*}

The fact that $I_{\textbf{A}}$ is not a Delta function of $k'$ can be utilized to quantify the total transition rate from the initial state $\mid\psi_{si_b}>$ to all valid final states. The total number of valence band electron states in the first Brillouin zone is twice the number of primitive cells in the sub-system, i.e., $2a_{z_ns}^{-3}V_s$ with $a_{z_ns}$ being the lattice constant of ZnS. Those are the potential states to be converted to holes due to photon absorption. The first Brillouin zone is a cube in k space with volume $8\pi^3/a_{z_ns}^3$. The corresponding $k'$ value can be calculated from each final state. The number of final states that fall in the range between $k'$ and $k'+dk'$ equals $4\pi k'^2{(2\pi)}^{-3}a_{z_ns}^3a_{z_ns}^{-3}V_sdk'$ or ${(2\pi^2)}^{-1}k'^2V_sdk'$. The factor $2$ is dropped because the photon-electron interaction $\frac{e}{2m_e}(\hat{\textbf{A}}\cdot\hat{\textbf{p}}+\hat{\textbf{p}}\cdot\hat{\textbf{A}})$ does not affect electron spin. Those states add to the transition rate by ${(2\pi^2)}^{-1}k'^2V_s\kappa_{if}dk'$. The lower limit of $k'$ is zero. While the exact upper limit $k'_{max}$ depends on $\mid\psi_{si_b}>$, because $\kappa_{if}$ drops quickly after $k'$ exceeds $1/(2c\tau)$, $k'_{max}$ can be assumed $\infty$. The total contribution $\kappa_{i_b}$ between $0$ and $k'_{max}$ is an integral
\begin{linenomath*}
\begin{equation}
\kappa_{i_b} = \int \limits_{0}^{\infty}{(2\pi^2)}^{-1}k'^2V_s\kappa_{if} \, dk'  \label{eq:52}
\end{equation}
\end{linenomath*}
which can be calculated as
\begin{linenomath*}
\begin{equation}
\kappa_{i_b} \approx \frac{4\alpha\omega_{if}}{c^2}\frac{E_p}{6m_e}\frac{n_k^3}{b_k^3}(1-\gamma)^2 . \label{eq:53}
\end{equation}
\end{linenomath*}
The right hand side of Eq.~\eqref{eq:53} equals the transition rate due to a plane wave vector potential and exact lattice momentum selection rule multiplied with $\frac{n_k^3}{b_k^3}(1-\gamma)^2$.     

The energy $E_{v\textbf{k}}$ of a hole with wave vector $\textbf{k}$ equals $E_v-\hbar^2\textbf{k}^2/(2m_e)$ where $E_v$ is the energy at the valence band edge \cite{18}. For an electron with effective mass $m^*_e$ in the conduction band, its energy $E_{c\textbf{k}}$ equals $E_c+\hbar^2\textbf{k}^2/(2m^*_e)$. The photon energy $\hbar\omega_{if}$ to enable the transition from the valence band to the conduction band is 
\begin{linenomath*}
\begin{equation}
\hbar\omega_{if} = E_c-E_v+\hbar^2\textbf{k}^2(\frac{1}{2m^*_e}+\frac{1}{2m_h}) . \label{eq:54}
\end{equation}
\end{linenomath*}
The photon energy $\hbar\omega_{if}$ equals the energy difference between the initial and final bound eigenstates of the incident electron. It is the result of spontaneous emission. The initial energy of the electron, however, may be much larger than this energy difference. More photons may be emitted to conserve the total energy. Consider a situation where multiple photons are emitted with the same vector potential. Their vector potential will add up. For a total of $n_{ph}$ photons, the magnitude of the accumulative vector potential will be $n_{ph}$ times that of the single photon. Since the transition rate as expressed by Eq.~\eqref{eq:40} is proportional to the absolute square of the vector potential, the transition rate will be $n_{ph}^2$ times as large as the original one due to a single photon emission.    

Another factor to consider is the impact of the relative permittivity $\epsilon_r$ to the transition rate. In the expression of Eq.~\eqref{eq:53}, both $\alpha$ and $c$ change with $\epsilon_r$. The term $\alpha/c^2$ in a media with relative permittivity $\epsilon_r$ will be $\sqrt{\epsilon_r}$ times that in free space.    

With the above generalization, the transition rate $\kappa_{\downarrow}$ from $\mid\psi_{si_b}>$ to $\mid\psi_{sf_b}>$ becomes
\begin{linenomath*}
\begin{equation}
\kappa_{\downarrow} = \sqrt{\epsilon_r}n_{ph}^2\frac{4\alpha\omega_{if}}{c^2}\frac{E_p}{6m_e}\frac{n_k^3}{b_k^3}(1-\gamma)^2 . \label{eq:55}
\end{equation}
\end{linenomath*}
This rate applies to all bound conduction state components of the incident electron.  
 
The incident electron may also transition to a bound eigenstate of an atom in the sub-system. Comparing this type of transition to the quantified transition, the energy of each emitted photon will be larger while the number of emitted photons will be smaller, and the averaged distance between the initial and end wave functions will be larger. Those differences may result in much lower transition rate, so quantification of the transition rate is omitted in this paper.     

\subsection{Stimulated transition rate from valence band to conduction band}\label{subsec:4.3}

It is clear from Eq.~\eqref{eq:54} that the energy of each emitted photon is larger than the band gap $E_g$, which equals $E_c-E_v$, so the photons will cause electron transitions from the valence band to the conduction band.  To quantify the transition rate for any specific state of the photons, the density function $\rho(E)$ in Eq.~\eqref{eq:40} does not apply to photons anymore. Instead, it changes to be the density of the state pairs that is based on the lattice momentum selection rule during the transition of electrons from the valence band to the conduction band. 

The transition rate Eq.~\eqref{eq:40} still applies although the initial state $\mid\psi_{si_b}>$ in this case is a valence state. Again, the only component that contributes to the transition is $\frac{e}{m_e}<u_{sf_b}\mid\textbf{A}_{sum}\cdot\hat{\textbf{p}}\mid u_{si_{b}}>$ or equivalently $\frac{e\mid I_{\textbf{A}}\mid}{m_e}<u_{sf_b}\mid\textbf{n}\cdot\hat{\textbf{p}}\mid u_{si_b}>$. Also $\textbf{k}_{f_b}$ and $\textbf{k}_{i_b}$ will approximate to each other. 
   
Combine the two terms $\frac{1}{2m^*_e}$ and $\frac{1}{2m_h}$ in Eq.~\eqref{eq:54} by using the reduced mass $m_{eh}$ or $(\frac{1}{m^*_e}+\frac{1}{m_h})^{-1}$. Since Eq.~\eqref{eq:54} is very similar to the dispersion relation of electrons in the conduction band, the density function $\rho(\hbar\omega_{if})$ can be obtained similarly, which is
\begin{linenomath*}
\begin{equation}
\rho(\hbar\omega_{if}) = \frac{1}{2\pi^2}(\frac{2m_{eh}}{\hbar^2})^{3/2}\sqrt{\hbar\omega_{if}-E_g} . \label{eq:56}
\end{equation}
\end{linenomath*}
Ignoring the very low percentage of occupied conduction states, then $\rho(\hbar\omega_{if})$ is the density of electron-hole pair states. The new transition rate $\kappa_{\uparrow}$ from $\mid\psi_{si_b}>$ to $\mid\psi_{sf_b}>$ can be obtained from Eq.~\eqref{eq:52} and Eq.~\eqref{eq:50} under the conditions that the last fraction on its right hand side being replaced with the right hand side of Eq.~\eqref{eq:56} and $\epsilon_0$ being replaced with $\epsilon_r\epsilon_0$. The result is
\begin{linenomath*}
\begin{equation}
\kappa_{\uparrow} = \frac{4\alpha\omega_{if}}{\epsilon_rc^2}\frac{E_p}{6m_e}\frac{({2m_{eh}c^2})^{3/2}}{2({\hbar\omega_{if}})^2}\sqrt{\hbar\omega_{if}-E_c+E_v}\frac{n_k^3}{b_k^3}(1-\gamma)^2  . \label{eq:57}
\end{equation}
\end{linenomath*}

Eq.~\eqref{eq:57} is the transition rate due to the impact of a single photon. Since there are multiple photons initially, the transition rate will be much higher at the beginning. Again the transition rate is proportionally to the square of the photon number. As more photons are absorbed, the transition rate will reduce till it reaches the lowest value as described by Eq.~\eqref{eq:57}. 

\subsection{Determination of the sub-system}\label{subsec:4.4}

Photons are emitted or absorbed randomly. It is the result of photon-charge interaction. Before an emission is realized, the energy of emitted photon is unpredictable. Since absorption is the reverse of emission, it is reasonable to assume that the absorbed photon is unpredictable either, which may be caused by the fluctuation of electromagnetic field in space. Thus the process of single photon emission or absorption can be viewed as the emission or absorption of one photon from an ensemble of candidate photons that are similar to each other. 

According to Eq. ~\eqref{eq:55} and Eq.~\eqref{eq:57}, the transition rates depend on the constituent of the sub-system. The contribution of the sub-system to the transition rates comes from $(1-\gamma)^2/b_k^3$. For any realized emission or absorption, the sub-system can be uniquely determined by maximizing the transition rate. This is equivalent to finding the unique positive value of $b_k$ such that the derivative of $(1-\gamma)^2/b_k^3$ with respect of $b_k$ equals zero, i.e., 
\begin{linenomath*}
\begin{equation}
2b_k^3e^{-b_k} = 6-3((b_k+1)^2+1)e^{-b_k} . \label{eq:58}
\end{equation}
\end{linenomath*}
   
For any atom, it is well known that the Hamiltonian of the atom is the summation of the Hamiltonian of relative motion and the Hamiltonian of the center-of-mass motion. The former contains all the electrons for multi-electron atoms. For hydrogen atoms, there is only one sub-system of relative motion. For multi-electron atoms, there are multiple choices of sub-systems of relative motion. It is well accepted that the atomic spectrum is the result of transitions between eigenstates of the relative motion of the sub-system that contains the relative motion of all the electrons. For any realized photon emission or absorption, that sub-system maximizes the transition rate. To see this, consider a different sub-system that leaves out one or more electrons. Since the electrons in the surroundings strongly interact with the sub-system, the eigenvalues of Eq.~\eqref{eq:2} changes rapidly with time. On the other hand, due to energy conservation, the energy of potential photon emission or absorption can only be the gap between two eigenvalues. Because the gap varies rapidly with time, before the process to emit  or absorb a photon for one target gap completes, that target becomes obsolete already. So the emission or absorption rate will be extremely low if not impossible at all.

\subsection{Numerical results}\label{subsec:4.5}

Some basic parameters of the detector material are needed for the calculation. For commonly used ZnS phosphor, the measured electron affinity of bulk ZnS material is between 3.8eV and 3.9eV \cite{19}. An affinity of $3.8eV$ will be used in this paper. The bandgap energy is $3.6eV$ \cite{20} and the optical frequency permittivity $\epsilon_r$ is $5.1$  \cite{21}. The widely cited work function of ZnS is $5.4eV$ \cite{22}. The maximum penetration depth of the $50keV$ incident electrons in ZnS was not provided in \cite{8}, but according to a different experiment \cite{23}, that value is $6um$. The effective electron mass $m^*_e$ of ZnS is $0.28m_e$ while the effective hole mass $m_h$ can be $0.18m_e$ and $0.86m_e$ \cite{24}. From these values, the reduced mass $m_{eh}$ is calculated to be $0.182m_e$. The value of $E_p$ for ZnS is unknown. Using the $E_p$ values of bulk III-V semiconductor \cite{17} as reference, since most of them are greater than $20eV$, $E_p$ of ZnS is assumed to be $20eV$. 

The average energy loss for $50keV$ electron is $8.3eV$ per $nm$. Using this constant rate and the fact that the length of the electron wave packet was about $1um$, assuming uniform probability distribution along the $1um$ length, then the energy of the wave packet was $4167eV$ when the wavefront reached the maximum penetration depth of $6um$. From that point on, it takes about $12.6$ femto-seconds for the end of the wave packet to travel $0.5um$. The total energy at that point is $2083eV$.

When the kinetic energy of an electron is less than $3.8eV$, its total energy becomes negative. The electron continues to loss kinetic energy when it propagates further into the detector. The energy loss per $nm$ varies with energy. For very low energy electron, the energy loss per $nm$ can be lower than $0.1eV$. There is no ZnS data available. Using extrapolated silicon data \cite{25} as an example, the energy loss rate when the energy is about $2.7eV$ above the Fermi level is only $0.1eV$ per $nm$, in comparison to about $0.5eV$ for $5.0eV$ energy. The $2.7eV$ energy corresponds to $1.1eV$ above $E_c$ since ZnS has $5.4eV$ work function. In a simple model, the wave packet has linear energy distribution between the wavefront and the end of the packet, the electron at the end of the packet loses $8.3eV$ for every $nm$ it travels, and the probability for the incident electron to be in the wavefront linearly increases with time. The electron energy in the wavefront is assumed to be $Ec+1.1eV$. Using this simple model, if a transition happens when the total packet energy is $2083eV$, a single photon will have $5.8eV$ energy and the maximum number of emitted photons $n_{ph}$ will be $359$. 

The numerical solution of Eq.~\eqref{eq:58} yields $b_k \approx 2.32$. According to Eq.~\eqref{eq:55}, the transition rate $\kappa\downarrow$ changes with time due to its time dependent factor $n_{ph}$. As time elapses, $n_{ph}$ reduces so does the transition rate. When $n_{ph}$ equals $359$, the transition rate $\kappa\downarrow$ is calculated to be $3.56e14$. The corresponding lifetime $\tau$ is $2.8$ femto-seconds and the radius $2b_kc\tau/\sqrt{\epsilon_r}$ is $1.7um$. The effective transition rate should be higher. Comparing the lifetime to $12.6$ femto-seconds, the chance to transition at that moment is almost $100\%$ times $P_{V_s}$, the total probability of the incident electron in all bound conduction states in $V_s$. If a transition occurs, it will emit $359$ photons. According to Eq.~\eqref{eq:57}, the absorption rate of one photon $\kappa\uparrow$ is calculated to be $9.5e14$. That means $1.1$ femto-seconds lifetime. The photons will travel inside the detector till they are absorbed. The traveling distance is calculated as $0.15um$. The $359$ photons will excite $359$ valence electrons to the conduction band.  
 
Using this example calculation, the incident electron has $50\%$ chance to transition and excite $359$ electrons to the conduction band. In case of ZnS with doped copper atoms, those electrons will fall back to the copper energy level and emit green light. The green color photons are the results of recombination of electrons and holes that are initially distributed in a sphere with $1.7um$ radius. The density of emitted photons in a region is proportional to the total probability of the incident electron in all bound conduction states in that region. The emitted photons are perceived as position measurement results. As long as the size of the emitted photons is much smaller than the $1.2mm$ interference pattern cycle, the resolution of position measurement results will be very high. 

There is also $50\%$ chance that the complement of the transitions will happen, i.e., the wave function at the end of the transitions will not contain any bound conduction states of the incident electrons. After such transition, the incident electron in the new wave function will start to lose energy again while it travels further into the detector. Sometime later transitions with some probability will happen to generate fewer than $359$ photons. So normalized to $100\%$ electron detection, the number of emitted photons will exceed $179$.    

The measurement recorded about $500$ photons on average. The example calculation is only $36\%$ of the measurement results. The percentage will be higher if other smaller number of photon emission events are included. But the ratio will no reach $72\%$. Yet, given the complexity of the problem and the simplicity of the linear model, the calculation-to-measurement comparison is acceptable. 

If the magnitude of the total vector potential of $359$ photons is used, then $\kappa\uparrow$ will be significantly higher than $\kappa\downarrow$ , meaning that the hole generation process is not a bottleneck for the spontaneous emission rate $\kappa\downarrow$. 

\section{Discussion}\label{sec:5}

\subsection{Photon confinement due to other broadening schemes}\label{subsec:5.1}

A fluorescent film was used in measuring the electron double-slit interference pattern \cite{8}. Typical transition time of fluorescent material after a charge carrier is released from a trap is on the order of $ns$. The natural spectral line broadening of this type of transition will result in photon packets with radius $2n_kc\tau/\sqrt{\epsilon_r}$ on the order of one meter. The region where photons are generated can be determined from the emitted photons. However, this process is similar to that in micro-light-emitting-diode (LED), i.e., photons are generated due to electron-hole recombination. The latter has achieved high resolution display with $5um$ LED pitch \cite{26}, meaning that the generated photons were within spheres of $2.5um$ radius. This contradicts the large photon size due to natural spectral line broadening. To resolve this issue, other broadening schemes need to be considered. But for now, it is safe to assume that the photon size is much smaller than the $1.2mm$ interference pattern cycle. 

\subsection{Generality and consistence}\label{subsec:5.2}

The analysis of the position measurement not only confirms the perception that a quantum measurement can happen at any location and any time but also showcases the generality and consistence of the new measurement theory. In principle, the charge-photon interaction used in quantifying the position measurement is the same as that used to calculate the lifetime of hydrogen eigenstates. The derived formulae for transition rates reveal the existence of maximum transition rate due to optimal sub-system. Conversely, the discovered principle that photon-charge interaction selects the most favorable sub-system to achieve the maximum transition rate also holds true for atoms. The surprise is that the appropriate sub-system of a multi-electron atom is well known but the reason for that choice has not been thoroughly investigated. Besides these, the theorem of the existence of the preferred basis is valid for any sub-system that interacts with its surroundings. 

For any measurement, one object can only be in one sub-system or the surroundings of a sub-system. So the optimal sub-systems selected by photon-charge interactions are separated from each other, i.e., they do not have any common object.       

\subsection{Future work}\label{subsec:5.3}

As one of the fundamental assumptions \cite{15} of quantum mechanics, the wave function of an object being measured becomes an eigenfunction immediately after a measurement. The transition process is unknown. That assumption applies to all the selected sub-systems too. The mechanism for a wave function that contains multiple eigenfunctions of a sub-system to a single component containing only one eigenfunction needs to be investigated.

\section{Conclusion}\label{sec:6}

A general and consistent measurement theory is presented in this paper with new findings. According to this theory, all measurements are the results of photon-charge interaction and are energy measurements. The analysis and numerical calculation provided in this paper clearly show that a commonly viewed position measurement was the result of energy measurement. The new findings provide mechanisms that uniquely determine the sub-system and the preferred basis for each measurement. With those mechanisms, the probability or rate for a measurement to occur is ultimately determined by the system Hamiltonian and wave function. The new theory solves the quantum measurement problem in the sense that its rigorous rules can be used to consistently answer when and where a measurement might occur and what state that measurement will end with.





\end{document}